# Single Aperture Large Telescope for Universe Studies (*SALTUS*): Science Overview

Gordon Chin[a*], Carrie M. Anderson[a], Jennifer Bergner[c], Nicolas Biver[d], Gordon L. Bjoraker[a], Thibault Cavalie[d], Michael DiSanti[a], Jian-Rong Gao[e], Paul Hartogh[f], Leon K. Harding[b], Qing Hu[g], Daewook Kim[h], Craig Kulesa[h], Gert de Lange[e], David T. Leisawitz[a], Rebecca C. Levy[h], Arthur Lichtenberger[i], Daniel P. Marrone[h], Joan Najita[j], Trent Newswander[k], George H. Rieke[h], Dimitra Rigopoulou[l], Peter Roefsema[e], Nathan X. Roth[a], Kamber Schwarz[m], Yancy Shirley[h], Justin Spilker[n], Antony A. Stark[o], Floris van der Tak[e], Yuzuru Takashima[h], Alexander Tielens[p], David J. Willner[o], Edward J. Wollack[a], Stephen Yates[e], Erick Young[q] Christopher K. Walker[h]

[a]NASA Goddard Space Flight Center, 8800 Greenbelt Road, Greenbelt, 20771, USA

[b] Northrop Grumman Space Systems, Dulles VA 20166, USA

[c]University of California, Berkeley, Berkeley, CA, 94720, USA

[d]LESIA CNRS, Observatoire de Paris 5 pl. J. Janssen 92190 Meudon, France

[e]Netherlands Institute for Space Research (SRON), Neils Bohrweg 4, 2333 CA Leiden, Netherlands

[f]Max Planck Institute for Solar System Research, Justus-von-Liebig-Weg 3, 37077 Göttingen, Germany

[g]MIT EECS, 77 Massachusetts Ave, Cambridge MA 02139, USA

[h]University of Arizona, Tucson, AZ, 85721, USA

[i]University of Virginia, Thornton Hall E-224 , Charlottesville, VA 22094, USA

[j]National Optical Astronomy Observatory (NOAO), 950 N. Cherry Ave, Tucson, AZ 85719, USA

[k]Space Dynamics Laboratory (SDL), 1695N. Research Parkway, Logan, UT 84341, USA

[l]Oxford University, Clarendon Laboratory, Parks Road, Oxford, OX1 3PU, UK

[m]Max Planck Institute for Astronomy, Königstuhl 17, 69117 Heidelberg, Germany

[n]Texas A&M University, 4242 TAMU, College Station TX 77843-4242, USA

[o]Center for Astrophysics Harvard Smithsonian, 60 Garden St., Cambridge, MA 92138, USA

[p]University of Maryland, 4296 Stadium Dr., College Park, MD 20742-242, USA

[q]Universities Space Research Association, 425 3rd Street SW, Suite 950, Washington DC, 20024, USA

**Abstract**. The *SALTUS* Probe mission will provide a powerful far-infrared (far-IR) pointed space observatory to explore our cosmic origins and the possibility of life elsewhere. The observatory employs an innovative deployable 14-m aperture, with a sunshield that will radiatively cool the off-axis primary to <45K. This cooled primary reflector works in tandem with cryogenic coherent and incoherent instruments that span the 34 to 660 *μ*m far-IR range at both high and moderate spectral resolutions. The mission architecture, using proven Northrop Grumman designs, provides visibility to the entire sky every six months with about ~35% of the sky observable at any one time. *SALTUS's* spectral range is unavailable to any existing ground or current space observatory. *SALTUS* will have 16x the collecting area and 4x the angular resolution of *Herschel* and is designed for a lifetime of ≥5 years. The *SALTUS* science team has proposed a Guaranteed Time Observations (GTO) program to demonstrate the observatory's capabilities and, at the same time, address high priority questions from the Decadal Survey[1] that align with NASA's Astrophysics Roadmap[2]. With a large aperture enabling high spatial resolution and sensitive instruments, *SALTUS* will offer >80% of its available observing time to Guest Observer (GO) programs, providing the science community with powerful capabilities to study the local and distant Universe with observations of 1000's of diverse targets such as distant and nearby galaxies, star forming regions, protoplanetary disks, and solar system objects.

**Keywords**: Far-infrared observatory, cooled deployable antenna, cryogenic heterodyne instrument, MKID grating spectrometer.

\* E-mail: gordon.chin@nasa.gov

# 1 Introduction

*Single Aperture Large Telescope for Universe Studies (SALTUS)* is a Probe class far-IR space observatory (**Fig. 1**) that will provide a new, powerful window to the Universe through which we can explore our cosmic origins. The unprecedented sensitivity and angular resolution of *SALTUS* is achieved by combining a passively cooled (<45K), 14-m off-axis telescope with an actively cooled instruments consisting of 1) a suite of heterodyne receivers (HiRX; see Silva et al., "The High Resolution Receiver (HiRX) for the Single Aperture Large Telescope for Universe Studies (SALTUS)," *J. Astron. Telesc. Instrum. Syst.* (this issue)) with resolving power R=$10^6$–$10^7$ and 2) a broad-band, 34–230 μm grating spectrometer (SAFARI-Lite; see Roelfsema et al., "The SAFARI-Lite Imaging Spectrometer for the SALTUS Space Observatory," *J. Astron. Telesc. Instrum. Syst.* (this issue)) with a resolving power of R~300. Instrument performances are summarized in **Table 1**. *SALTUS* can observe numerous atomic and molecular species, as well as lattice modes of ices and minerals which, due to atmospheric absorption and/or wavelength coverage, are beyond the capabilities of existing ground-based or space-based observatories.) (**Fig. 2**). *SALTUS* provides a set of capabilities that directly complements JWST and ALMA and greatly advances on previous far infrared missions (see **Figure 3)**.

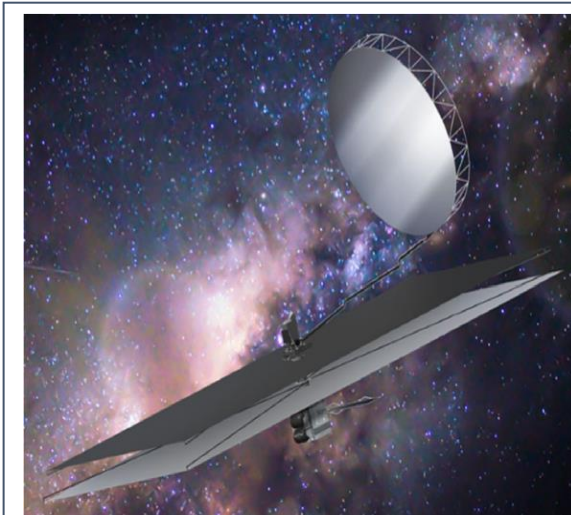

**Fig. 1** The SALTUS Observatory uses a radiatively cooled, inflatable 14-meter off-axis aperture with sensitive far-IR, high and moderate resolving power systems to open a new window on our Universe. See Sec 3 for details on the SALTUS observatory architecture

**Table 1** *SALTUS* Instruments performs observations over a wide wavelength range with both high resolution coherent and moderate resolution incoherent detectors at high spatial resolution.

| Instrument | Wavelength (μm) | Resolving Power | Sensitivity (5σ/1 hr) | Detector Type | Beam Size (arcsec) |
|---|---|---|---|---|---|
| SAFARI-Lite | 34 – 230 | 300 | ~ $2 \times 10^{-20}$ W/m$^2$ | MKID Arrays Incoherent | 0.6 - 4.1 |
| HiRX | 56 - 660 | $10^6 - 10^7$ | ~ 100 mK | SIS/HEB Mixers Coherent | 1 – 11.7 |

With its large collecting area and high sensitivity *SALTUS* is uniquely capable of advancing the Decadal Survey's [1] high-priority science themes of "*Cosmic Ecosystems*" and "*Worlds and Suns in Context*" The *SALTUS* Science Objectives are listed by theme in **Table 2** and map directly to numerous Decadal questions (shown in brackets) and the NASA Astrophysics Roadmap [2]. SALTUS's capabilities will enable significant progress towards resolving high-priority Decadal Survey questions in the areas of:

**Fig. 2** Simulated terrestrial atmospheric transmission spectrum (black) demonstrating SALTUS's far-IR spectral region, inaccessible from the ground, and outside of *JWST* (green) and *ALMA's* (magenta) operational wavelengths. The tunable HiRX Bands 1 – 4 with SAFARI-Lite (blue), target critically important wavelengths that are significantly or completely blocked from the ground; this includes low energy transitions of $H_2O$ and its isotopologues, and other species such as HD 1-0 and HD 2-1.

1. measuring the formation and buildup of galaxies, heavy elements, and interstellar dust from the first galaxies to today, probing the co-evolution of galaxies and their supermassive black holes across cosmic time, and,

2. tracing the astrochemical signatures of planet formation (within and outside the Solar System).

The *SALTUS* observatory's flexible capabilities provide the astrophysics community with a robust Guest Observer (GO) program during a baseline 5-year mission, which could be extended a factor of two or longer. All *SALTUS* observations will be archived and made available to the science community through Caltech/IPAC within 6 months. The archived *SALTUS* observations can be mined by a Guest Investigation (GI) program. The PI-led, Guaranteed Time Observations (GTO) discussed briefly in Sec. 4, are aimed to highlight *SALTUS's* capabilities, promising high impact science, with illustrative GO science objective examples offered in Sec. 5. More than 80% of the available observational time will be allocated for the GO program, which will foster unexpected explorations and a rich harvest of unforeseen discoveries (see Sec 6).

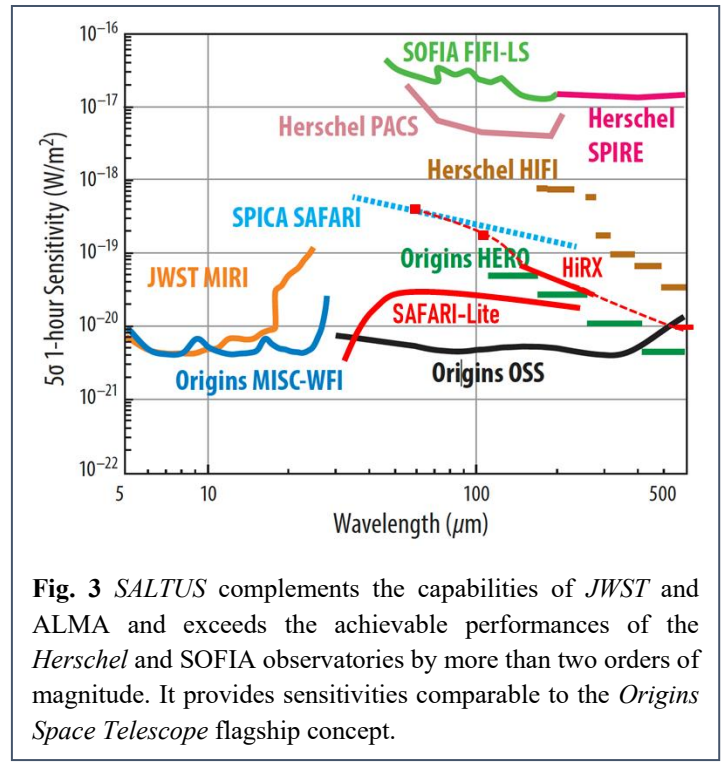

**Fig. 3** *SALTUS* complements the capabilities of *JWST* and ALMA and exceeds the achievable performances of the *Herschel* and SOFIA observatories by more than two orders of magnitude. It provides sensitivities comparable to the *Origins Space Telescope* flagship concept.

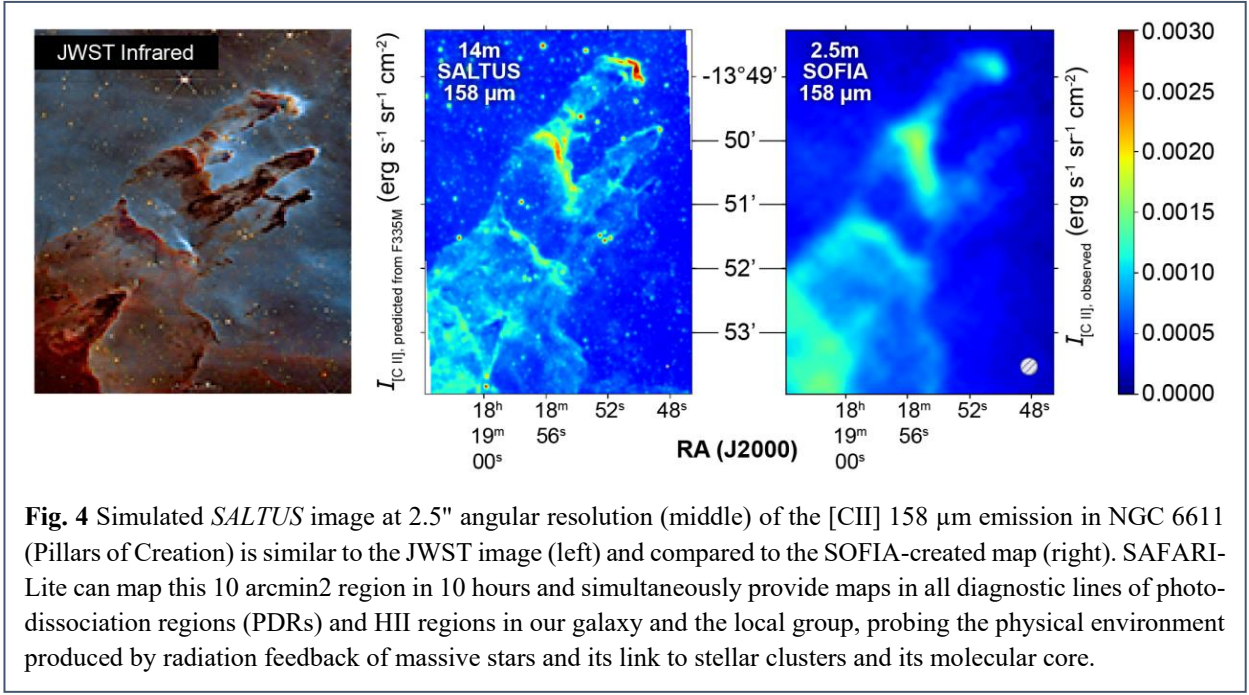

**Fig. 4** Simulated *SALTUS* image at 2.5" angular resolution (middle) of the [CII] 158 µm emission in NGC 6611 (Pillars of Creation) is similar to the JWST image (left) and compared to the SOFIA-created map (right). SAFARI-Lite can map this 10 arcmin2 region in 10 hours and simultaneously provide maps in all diagnostic lines of photo-dissociation regions (PDRs) and HII regions in our galaxy and the local group, probing the physical environment produced by radiation feedback of massive stars and its link to stellar clusters and its molecular core.

## 2   The Need for Aperture

*SALTUS* will be the first far-IR space observatory with a large enough aperture to provide arcsec-scale spatial resolution (0.66 – 10 arcsec) in the far-IR. This will permit an unmasking of the true nature of the cold Universe, which holds the answers to many of the questions concerning our cosmic origins. *SALTUS* will break through the confusion limit in extragalactic observations (see below) that plagues smaller apertures [3] and probe the far-IR Universe at unprecedented detail (**Fig. 4**). The *SALTUS* aperture allows it to attain sensitivity performance of order that of the proposed *Origins Space Telescope* and *31 times* greater than that of *SOFIA* and *16 times* greater than that of *Herschel* (Figure 4). This increase in sensitivity means *SALTUS* will be able to probe orders of magnitude deeper than any past far-IR mission. A large aperture also helps minimize the impact of beam dilution on spectral line observations when targeting objects smaller than the telescope's diffraction-limited beam.

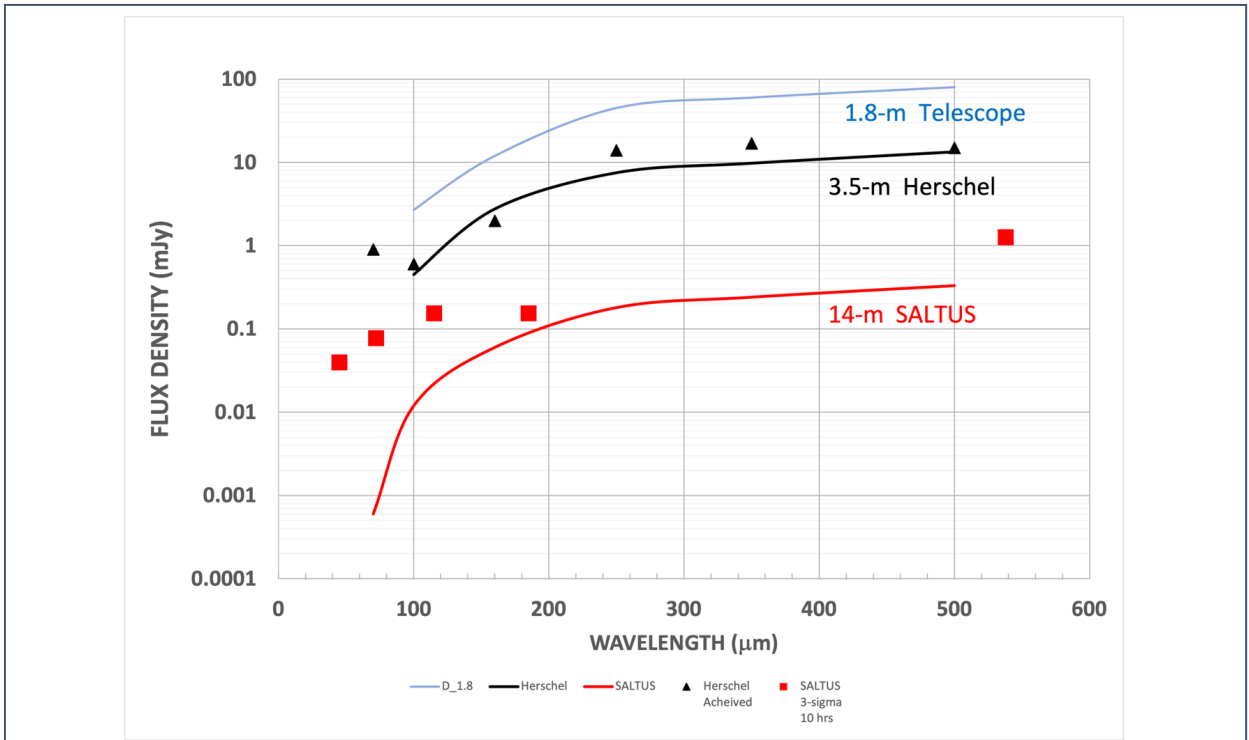

**Fig. 5** 3σ confusion limit curves in the far-IR as functions of wavelength for three different sized telescopes: a possible 1.8 m (blue), the 3.5 m *Herschel* (black), and the 14-m SALTUS (red). The individual points indicate the best achieved sensitivity for *Herschel* and the projected sensitivity for SALTUS mapping a square arcminute after only 10 hours of observation, demonstrating how SALTUS greatly outperforms smaller aperture space observatories.

Beam size and confusion noise are fundamental limits to the sensitivity and information content of far-IR observations. Confusion noise is defined to be the spatial variation of the sky intensity in the limit of infinite integration time and can only be directly addressed by increased telescope aperture size.

An observatory's minimum beam area is set by diffraction, which depends on the inverse square of the product of the aperture and wavelength. At some sensitivity level, the number of distant galaxies will exceed the number of beams, and the sensitivity cannot be improved with better instrumentation or longer integration times. The performance of current instrumentation indicates that aperture size will be the key driver in probing deeper. Using SPIRE on Herschel (3.5m), a confusion limit of 5.8, 6.3 and 6.8 mJy/beam at 250, 350 and 500 μm, respectively, was quickly reached [4]. This is illustrated in **Fig. 5**, showing the 3σ confusion noise curves for *SALTUS*, *Herschel*, and a possible two-meter class facility. We assume a differential source count as a function of flux density S of N(S) ~ $S^{-\gamma}$, with γ = 1.7, which is a reasonable value based on *Herschel* data. **Fig. 5** illustrates two key points. First, the 3.5m *Herschel* observatory had essentially reached the confusion limit and could not go deeper. Second, *the 14-m SALTUS observatory will break the 1 mJy* **confusion** *limit, necessary to understand the true nature of the distant far-IR universe, and greatly outperform past and anticipated space-borne facilities*. It can go much deeper than *Herschel* with only ten hours of observations assuming ***existing detector*** performance. *SALTUS will be greater than two orders of magnitude more sensitive than a cooled two-meter class facility*. The confusion limit is not just an issue for the cosmic far-IR background, but also strikes at the heart of many key questions; including the contribution of dust obscured star formation to the star formation history of the universe and the rise of the PAH/dust abundance with cosmic time.

Existing extragalactic continuum far infrared surveys are severely confusion limited. While low angular resolution surveys conducted with sufficiently high spectral resolution may address the confusion limit, they are biased by the emission from the brightest galaxies in a pixel. The contribution from fainter galaxies at a given wavelength will be easily lost in the glare of the continuum "headlight" of the dominant galaxy/ies. Spectral deconvolution techniques may be employed, but require the character of the galaxies in the field (together with their infrared spectrum) to be understood to a high degree of confidence, which will often not be the case. Indeed, this is the information we seek to measure. Studies performed for the *Origins* mission concept demonstrated that, as there are many sources per beam, spatial blending is a significant issue for narrow band continuum SED images – even for a 5.9m telescope – and simulations recover only 40% of the sources. Deblending techniques for emission line sources are slightly more effective as they recover about 50% of the sources in narrow band images [5]. Due its native angular resolution, SALTUS can avoid much of the issues related to the far-IR continuum glare of the distant universe that plague smaller telescopes and, therefore, does not require *a priori* knowledge or assumptions to interpret its observations. **Figure 6** is a simulated *SALTUS* multiband deep blind survey. Without the resolving power of SALTUS, the existence of the vast majority of low luminosity galaxies would be lost in the glare of the brighter objects, obscuring the true nature of the early universe. To understand galaxy evolution, it is imperative that we measure far-IR spectra of ordinary galaxies – not only exceptionally luminous galaxies - at redshifts out to the epoch of cosmic noon (i.e., $0 < z < 3$), and that requires sensitivity of the order of tens of micro-Jy, a level attainable with SAFARI-Lite on SALTUS, but not with a *Herschel*-sized or smaller telescope.

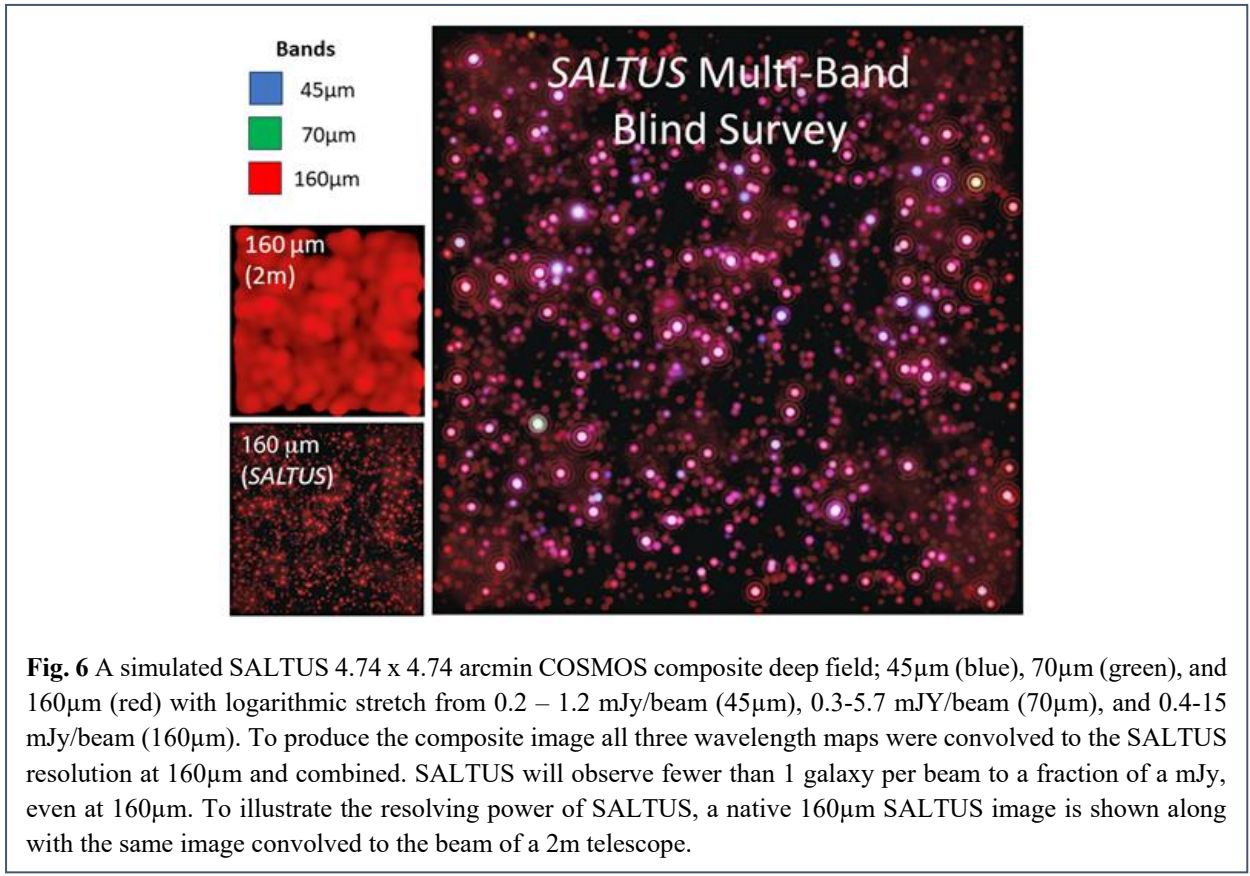

**Fig. 6** A simulated SALTUS 4.74 x 4.74 arcmin COSMOS composite deep field; 45µm (blue), 70µm (green), and 160µm (red) with logarithmic stretch from 0.2 – 1.2 mJy/beam (45µm), 0.3-5.7 mJY/beam (70µm), and 0.4-15 mJy/beam (160µm). To produce the composite image all three wavelength maps were convolved to the SALTUS resolution at 160µm and combined. SALTUS will observe fewer than 1 galaxy per beam to a fraction of a mJy, even at 160µm. To illustrate the resolving power of SALTUS, a native 160µm SALTUS image is shown along with the same image convolved to the beam of a 2m telescope.

## 3   *SALTUS* Observatory architecture

The *SALTUS* observatory (**Figure 1**) is based on the Northrop Grumman LEOStar-3 spacecraft product line. The observatory is comprised of the deployable telescope inflation control system, Sunshield Module (SM), Cold Corrector Module (CCM), Warm Instrument Electronics Module, and Primary Reflector Module (PRM). The primary reflector, M1, is a 14-m off-axis inflatable membrane radiatively cooled to <45 K by a ~1000 m$^2$ two-layer sunshield and is deployed via a high-heritage telescoping boom. The primary reflector, M1, manufactured by L'Garde, is new to astrophysics but has been employed by the telecom industry for decades. *SALTUS* will use a "seamless approach" in conjunction with thermo-formed gores. Seamlessness is achieved by using two sets of identical gores bonded in a staggered manner. The presence of the thicker seam tapes at the sealing line areas is designed into the initial uninflated configuration so that on inflation, M1 obtains a surface very close to an ideal paraboloid. The thermoforming of gores of the M1 membrane also results in the minimization of thickness variation over the entire surface. Using this approach there are no visible or measurable distortions at the seams, nor at the perimeter (see Arenberg et al. in this special volume). Before reaching the instruments the emergent beam from the primary reflector passes through the CCM (see Kim et al., "14-m aperture deployable off-axis far-IR space telescope design for SALTUS observatory," *J. Astron. Telesc. Instrum. Syst.* (this issue). The CCM corrects for residual aberration from the primary reflector and delivers a focused beam to the two SALTUS instruments – the High Resolution Receiver (HiRX) and SAFARI-*Lite*. The CCM and PRM reside atop a truss-based composite deck which also provides a platform for the attitude control system. The spacecraft bus, subsystems, and single-wing solar array are designed to withstand thermal backloading from the sunshield. The *SALTUS* observatory maintains a 90º attitude (±20º pitch; ±5º roll) off the primary reflector's boresight with respect to the Sun, facilitating the <45 K thermal environment for the primary mirror. *SALTUS* will reside in a Sun-Earth Halo L2 orbit with a maximum Earth range of 1.8 million km. The SALTUS observatory's instantaneous field of regard provides two continuous 20º viewing zones around the ecliptic poles resulting in full sky coverage in six months. Details of the *SALTUS* observatory architecture are presented in Harding et al., "SALTUS Probe Class Space Mission: Observatory Architecture and Mission Design," *J. Astron. Telesc. Instrum. Syst.* (this issue).

## 4 SALTUS Science Objectives

The PI-led GTO program addresses the *SALTUS* Science Objectives listed in **Table 2**. In Sec. 5 we provide examples of potential GO/GI programs enabled by *SALTUS* discovery space.

*4.1 SALTUS Theme 1: Cosmic Ecosystems*

*Theme 1* corresponds to the Decadal Survey's "*Cosmic Ecosystems*" theme and addresses the key science questions identified for a far-IR probe-class mission. As a versatile, sensitive

**Table 2** *SALTUS* Science Objectives Address Relevant Decadal Themes and Questions.

| Decadal Theme 1: Cosmic Ecosystems | Decadal Theme 2: Worlds and Suns |
|---|---|
| SALTUS **Science Objective 1:** **Trace galaxy and black hole co-evolution and heavy element production over cosmic time.** | SALTUS **Science Objective 2:** **Probe the physical structure of protoplanetary disks and follow the trail of water and organics from protoplanetary disks to the solar system.** |
| **1.1 What is the role of star formation in feedback in the Local Universe?** [F-Q2, D-Q2] | **2.1 How does the mass distribution in protoplanetary disks affect planet formation?** [F-Q4, F-Q4b, E-Q1c, E-Q3b] |
| **1.2 When did metals and dusts form in galaxies, affecting the process of star formation?** [D-Q2, D-Q4] | **2.2 What is the spatial distribution and evolution of water vapor and ice in protoplanetary disks?** [E-Q1c, F-Q4b, E-Q3b] |
| **1.3 What are the roles of feeding black holes in galaxies from the early Universe to today?** [D-Q3, D-Q4] | |
| **1.4 Which feedback mechanism dominates as a function of time over cosmic history?** [D-Q3] | **2.3 How did Earth and Ocean Worlds get their water?** [E-Q2c, E-Q3b] |

observing with arcsecond-level spatial resolution, *SALTUS* is the natural complement to the near- and mid-IR capabilities of *JWST*. This theme outlines an ambitious science campaign designed to answer outstanding questions raised by the Decadal Survey, including: *How do gas, metals, and dust flow into, through and out of galaxies? How do supermassive black holes form and how is*

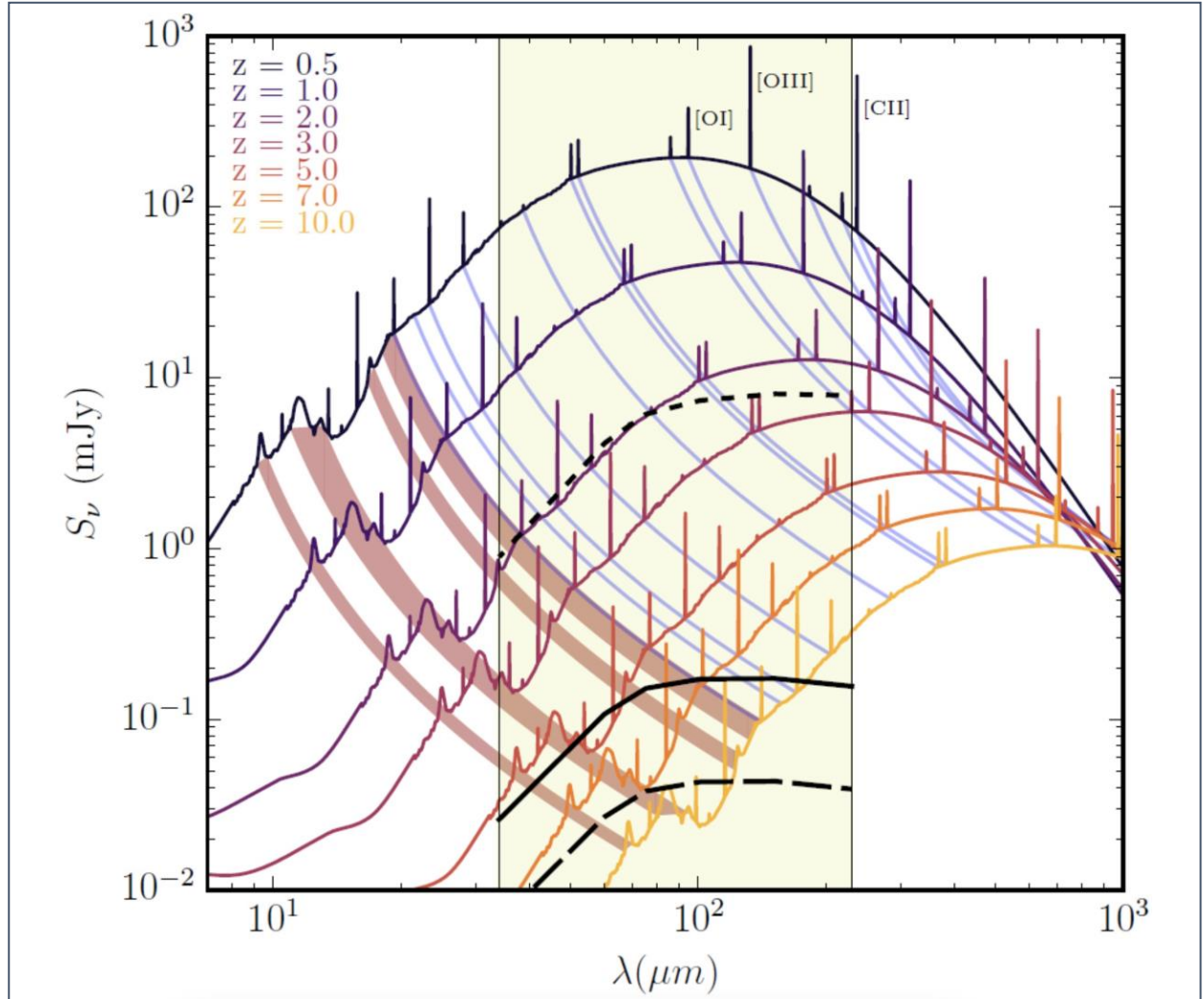

**Fig. 7** The spectral range of SALTUS over cosmic time. Schematic representation of the spectral energy distribution of a dusty $3\times10^{12}$ L$_\odot$ star forming galaxy with redshift. Lines important to the science case and PAH features are traced through redshift, and dominant cooling lines ([OI], [OIII], [CII]) are labeled. Out to z ~ 3, SAFARI-Lite probes the peak of the dust continuum and the bulk of the dust emission. Beyond z ~ 3, SAFARI-Lite takes over from JWST/MIRI to probe the red-shifted mid-IR PAH emission features. The yellow color-coded region indicates the wavelength range of SAFARI-Lite. The lower solid black curve is the detection limit for SAFARI-Lite at R = 300 for pointed observations (1 hour, 5σ). The lower long-dashed line approximates a detection limit for wide PAH features, which span many channels. The short-dashed line is the SAFARI-Lite detection limit in mapping mode (1arcmin$^2$ area mapped in 1-hr at 5σ).

*their growth coupled to the growth of their host galaxies? Theme 1* will make use of *SALTUS's* unique capabilities to address the SALTUS Science Objective 1: Trace galaxy and black hole co-evolution and heavy element production over cosmic time, which comprises four key science investigations, listed in **Table 2**, and expanded in Spilker et al., "Distant Galaxy Observations," *J. Astron. Telesc. Instrum. Syst. (this issue)*, and Levy et al., "Nearby Galaxy Observations," *J. Astron. Telesc. Instrum. Syst.* (this issue).

Thanks to the broad simultaneous wavelength coverage of the SAFARI-Lite, in particular see **Fig. 7**, extragalactic observations of individual targets, pointed surveys of well-defined samples, and blank-field spectral mapping campaigns are all capable of addressing different aspects of the *Cosmic Ecosystems* science goals. Many of these goals (and beyond) will also be addressed by the wide variety of community GO observing programs enabled by the versatile and sensitive *SALTUS* capabilities. Moreover, *SALTUS'* nearby galaxy GTO observing program centers on star-forming spirals and starburst galaxies, but we expect the versatility of *SALTUS* to inspire a wide range of community GO programs targeting galaxies in the nearby Universe. Such GO programs will greatly expand the range of physical conditions in which galactic feedback processes are characterized to include a range of galaxy mass, star formation rate, and metallicity compared to the GTO observing effort.

*4.2 SALTUS Theme 2: Worlds and Suns in Context*

The transformation of gas and dust clouds into stars and planetary systems is arguably the most important of astrophysical processes (**Fig. 8**). Star formation involves the full gamut of physical processes and a vast dynamic range in density and size scales. Much progress has been made in our understanding, but there are many gaps in our knowledge. The discovery of thousands of new planetary systems in recent decades has emphasized the amazing range and diversity of outcomes when building systems. *What are the conditions that lead to these vastly different outcomes?*

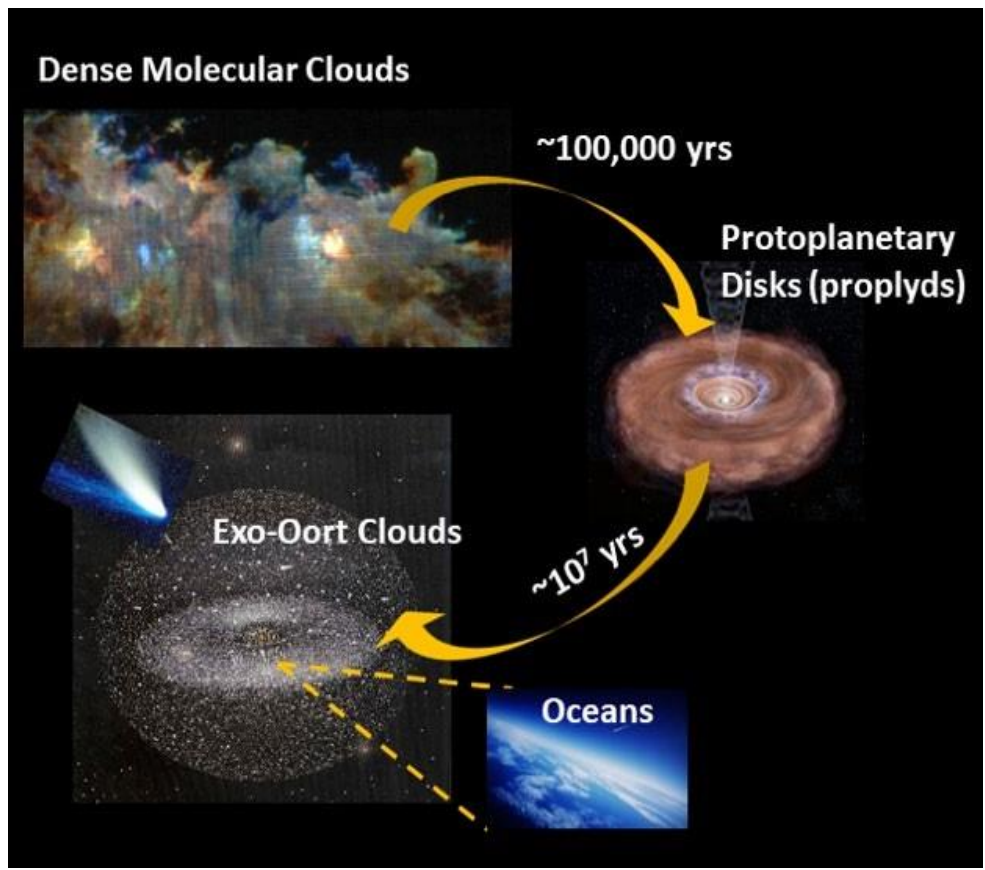

**Fig. 8** *SALTUS* follows the water trail from molecular clouds to oceans. The habitability of planets is closely tied to the presence of $H_2O$, which is formed in the shielded interiors of molecular clouds, transported to planet forming disks where volatiles are further chemically processed before becoming part of planetesimals and comets beyond the snow line. Planetesimals and comets then deliver these volatiles to terrestrial planets and ocean worlds. *SALTUS* is designed to probe this important journey using low lying rotational $H_2O$ lines that probe cold gas with HiRX and the icy grain reservoir through their phonon modes in emission with SAFARI-Lite while we expect GOs to probe the later stages.

The aim of Theme 2 is to study the structure and evolution of protoplanetary systems over a wide range of conditions, and to understand the origin of Earth's water. This theme defines a science campaign to answer several Decadal Survey questions [1], and enable significant progress towards resolving fundamental questions, such as: *How do habitable environments arise and evolve within planetary systems? Where does Earth's water come from? How did we get here?*

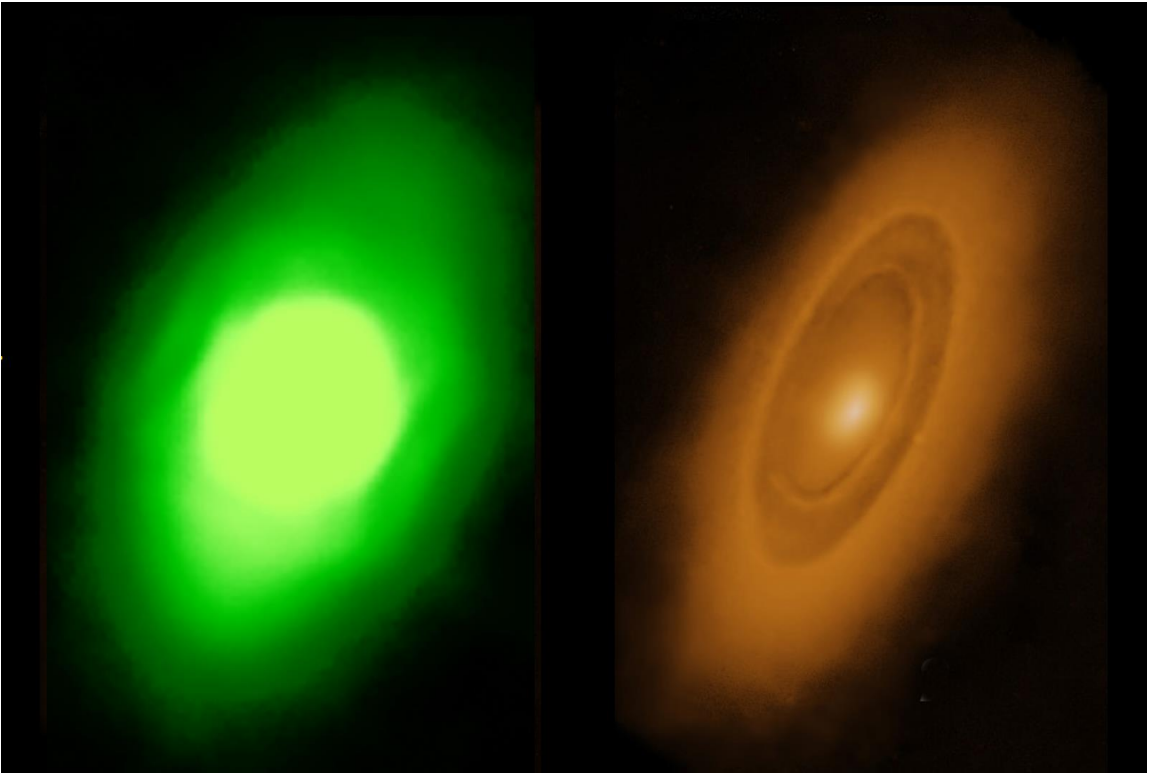

**Fig. 9** Image of the Fomalhaut planetary debris system with Spitzer at 24 µm (left) and JWST at 25.5 µm (right). An increase in telescope aperture from by a factor of seven (e.g. from 2 meters to 14 meters in the far infrared) yields a huge increase in information content of the images.

*SALTUS's* large aperture provides high sensitivity and spatial resolution that is coupled with the high spectral resolution of the HiRX instrument and the broad spectral coverage of the SAFARI-Lite instrument, this allows *SALTUS* to address Science Objective 2: *Probe the physical structure of protoplanetary disks and follow the trail of water and organics from protoplanetary disks to the solar system*. Figure 9 shows the revolutionary impact of the SALTUS angular resolution on a protoplanetary disk in the far-IR.

*SALTUS* observations will obtain the disk gas masses in hundreds of protoplanetary systems without the need for ancillary data. In a subset of the brightest disks, *SALTUS* HiRX will spectrally resolve strong HD lines at ~ 1 km/s velocity resolution. Because disk rotation follows a Keplerian velocity profile, the radius at which gas emission originates can be determined from the line profile. Thus, high spectral resolution observations of molecular lines in disks can be used to determine the radial location of the emission without having to spatially resolve the disk; a technique known as Doppler tomography or tomographic mapping. The spectral resolution of *SALTUS* HiRX is < 1 km s$^{-1}$, sufficient to distinguish emission originating in the inner versus outer disk. With its 3.5-meter aperture, Herschel was only able to detect a handful of protoplanetary disks in HD. A far-IR telescope with a significantly larger aperture than Herschel is *essential* to the success of such a study.

*SALTUS* is also well suited to answer a number of open questions in planetary science, including; how did the Earth and Ocean Worlds get their water? SALTUS will address this by measuring and comparing the HDO/H$_2$O ratio in a large sample of cometary reservoirs and planetary bodies (**Fig. 10**). This and other planned GTO science investigations are expanded on in

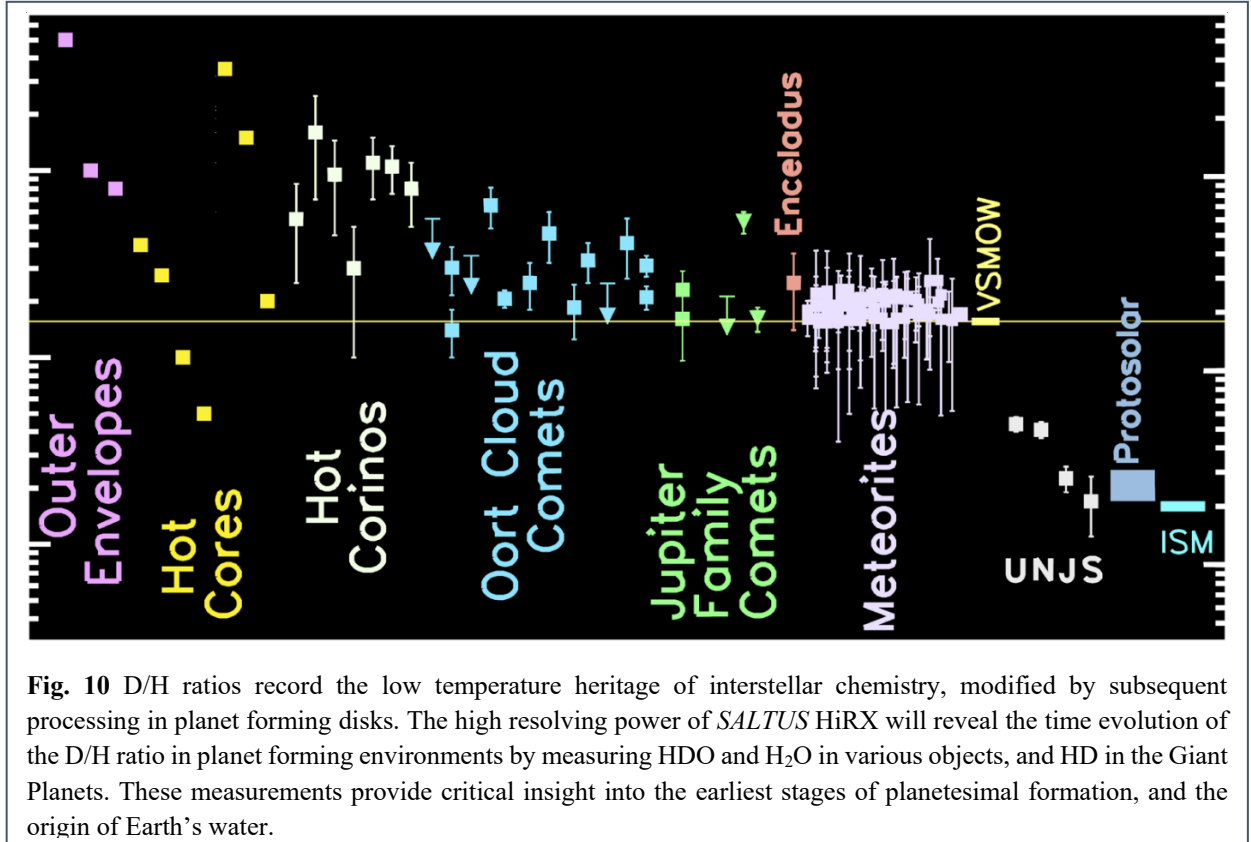

**Fig. 10** D/H ratios record the low temperature heritage of interstellar chemistry, modified by subsequent processing in planet forming disks. The high resolving power of *SALTUS* HiRX will reveal the time evolution of the D/H ratio in planet forming environments by measuring HDO and H$_2$O in various objects, and HD in the Giant Planets. These measurements provide critical insight into the earliest stages of planetesimal formation, and the origin of Earth's water.

Schwarz et al., "Star and Planets," *J. Astron. Telesc. Instrum. Syst.* and Anderson et al., "Solar System," *J. Astron. Telesc. Instrum. Syst.* within this Special Section.

*4.3 Programmatic Value of SALTUS Science*

*SALTUS* measurements extend the pioneering efforts of *Herschel*, enhance the scientific return of ground and space-borne observatories such as *ALMA* and *JWST*, complement missions such as *GUSTO*, *SPHEREx*, and *Roman*, and resonate with the science visions of ESA's *SPICA* and NASA's *Origins Space Telescope* mission concepts. As indicated in Tables 2, SALTUS addresses crucial, high-priority scientific inquiries raised in the Decadal Survey.

**5 Guest Observer /Guest Investigator Potential**

The potential GO Science programs listed below were constructed to demonstrate the science capabilities of *SALTUS* beyond the GTO observational agenda. As such, these GO Science Investigations are by no means exhaustive, and we fully expect they will be modified during the GO program selection. Enabled by the large discovery space in sensitivity, spatial resolution, spectral coverage, and mapping speed, *SALTUS* is truly a community resource, and there are a significant number of high-impact investigations possible.

*5.1 Debris disk architecture and search for true Kuiper Belt Analogues*

The debris disk phase follows the protoplanetary disk phase. Debris disks are gas-poor, with rings of second-generation dust thought to be influenced by the presence of planets; a ring-sculpting planet has been confirmed in at least one system. Debris disk observations allow us to study populations of small bodies around other stars and infer the presence of planets that otherwise evade detection. Debris disks also provide insight into the composition of solid bodies in other planetary systems. The detection frequency of debris disks decreases with system age. As the frequency of collisions drops, less observable dust is produced, thus, existing detections are biased toward young systems and the warmer dust around A and F stars [5]. Emission from true Kuiper Belt analogs, debris disks with the same intrinsic luminosity as the Solar System's Kuiper Belt, has yet to be observed because they are much too faint. These exo-Kuiper Belts have typical temperatures of 50K, corresponding to a black-body emission peak in the far-IR. Updating sensitivity estimates from the original *SPICA* SAFARI [6]. to *SALTUS's* SAFARI-Lite, *SALTUS* will reach the 5σ sensitivity threshold to detect exo-Kuiper belts around the nearest 30 G and K stars with known debris disks in 1 hour of integration. *SALTUS* can determine the frequency of exo-Kuiper Belts, characterizing how common dust is in mature planetary systems, thus addressing Decadal Question E-Q1d. Additionally, the angular resolution of *SALTUS* should enable mapping of the dust in some of these systems (Fig. 10).

*5.2 C/O Abundances at Late Stages*

*SALTUS* can observe young debris disks, expected to contain detectable levels of carbon and oxygen; based on the gas production model developed by [7,8]. These observations, focusing on ionized carbon and neutral oxygen, will complement those made by *ALMA*, which targets CO and neutral carbon [9, 10]. By doing so, *SALTUS* can gather valuable information about the carbon ionization fraction, a crucial factor in understanding the dynamics of gas, e.g., how well gas is coupled to the stellar wind. *SALTUS's* high spectral resolution (sub-km/s) can access spatial information such as the gas density and its position compared to dust. The ionization fraction and spatial information are essential for studying the transport of angular momentum in these disks and to determine the electron density that influences non-LTE excitation. Since gas is observed to be spreading from its original source (the planetesimals), it is important to understand if it is due to the magneto-rotational instability as discussed by [11], or MHD winds or even some hydrodynamic instabilities such as VSI or RWI [11].

The low surface density in debris disks allows a high photon flux from the central star, converting molecules like CO, $CO_2$, and $H_2O$ into CII and OI via photodissociation and photoionization. By targeting CII and OI in the far-IR, *SALTUS* users will gain insights into the initial species released from planetesimals by examining the C/O ratio, e.g., investigate whether CO, $CO_2$, or $H_2O$ is released. These observations will provide a comprehensive understanding of the gas disk composition at different radii from the central star. This knowledge is fundamentally important since there are indications that the late gas (detected in systems as old as 0.5 Gyr) can expand inward due to viscosity and reach the planets. Already formed planets can then accrete this gas at a high rate, thereby altering their initial atmospheric compositions. The accretion of carbon and oxygen by young planets may play a pivotal role in the formation of the building blocks of life or affect the temperature through greenhouse effects, thus influencing its habitability.

Currently, debris disk studies have been mainly with A stars. *SALTUS* has the sensitivity to detect CII and OI in the more typical FGK stars. These observations will determine the C/O ratio across spectral type during the late stages of planet formation, when volatile gasses are delivered to terrestrial planets, and addresses Decadal question E-Q3a "How are potentially habitable environments formed?" *SALTUS* particularly can look at this question around solar mass stars.

*5.3 Survey of water vapor in planetary atmospheres*

With HiRX Bands 1 and 2 *SALTUS* will measure $H_2O$ abundances in the stratospheres of the Gas Giants to determine whether planetary rings, icy moons, interplanetary dust particles, or comet impacts deliver water to these planets. Titan, another Ocean World, and the largest satellite of Saturn, has a complex atmosphere containing oxygen compounds that may have been delivered from the H$_2$O torus and, ultimately, from Enceladus. *SALTUS* will measure the vertical and horizontal distributions of H$_2$O in Jupiter and Saturn's stratospheres to constrain its external source. Gaseous H$_2$O emission has also been observed from the dwarf planet, Ceres; however, its source and spatial distribution are not known. The high sensitivity of *SALTUS* will permit the determination of whether cryo-volcanism or ice sublimation from localized regions is the source of water. This addresses Decadal Question E-Q2c.

*5.4 Stellar feedback in regions of massive star formation*

This topic aims to understand the role of stellar feedback in regions of massive star formation: How large is the kinetic energy input into the ISM, how does this depend on the characteristics of the region, and, connected to this, how does the surrounding medium react? This objective is addressed by spectral-spatial mapping of regions of massive star formation in the dominant far-IR cooling lines of PDR gas using both the SAFARI-Lite and HiRX instruments (c.f., the Pillars of Creation in Fig 4). The sample consists of 25 regions with typical sizes of 25 square arc minutes, covering a range in stellar cluster properties (single stars to small clusters to superstar clusters), GLIMPSE bubble morphology (ring-like, bi- or multi-polar, complex), environment (isolated star forming region, galactic mini starburst, nuclear starburst), and cluster age (0.1 to 5Myr).

SAFARI-Lite will provide full spectra over the 34-230 $\mu$m range (see Table 1), containing the key diagnostic atomic fine-structure transitions e.g., [OI] 63, 145 $\mu$m, [CII] 158 $\mu$m, [SiII] 34.5 $\mu$m, [OIII] 52, 88 $\mu$m, [NII], 122, 205 $\mu$m, and [NIII] 57$\mu$m, as well as high J CO transitions, which provide the physical conditions in the neutral PDR gas as well as the photo-ionized gas. The HiRX-2 will measure the dynamics of the PDR gas by mapping the [CII] 158$\mu$m line at km/s resolution and determine expansion speed and turbulent line width of the swept-up, expanding shell. *SOFIA*/upGREAT studies of the [CII] 158 μm emission of regions of massive star formation have demonstrated the feasibility of this technique [12]. Together these two data sets will quantify the kinetic energy of the expanding shell as well as the thermal, turbulent, and radiation pressures on the shell that can be directly compared to radiative and mechanical energy inputs of the stellar cluster.

*5.5 Put early-Universe feedback in context with local analogs of the first galaxies*

The first year of *JWST* observations revealed that early generations of galaxies were typically much lower in mass and significantly less metal-enriched compared to the massive star-forming spiral galaxies that dominate star formation today. Consequently, their UV radiation fields were harder [13], which impacted the gas ionization state and cooling rate. The shallow gravitational potential wells also enhanced the effects of supernova-driven feedback since single supernovae could eject up to ~95% of the heavy elements formed during the star's lifetime into the circum-galactic medium [14, 15]. While *ALMA* now allows the key far-IR fine structure diagnostic lines (e.g., [CII], [OIII]) to be detected at high redshifts, these lines still take several hours to detect even in the most massive and UV-luminous reionization-era galaxies [16] and are generally out of reach for the more typical galaxies thought to drive cosmic reionization.

Thanks to community efforts, many nearby galaxy populations have been identified bearing similarities to UV-bright reionization-era galaxies in terms of their mass, metallicity, and harsh UV radiation fields. Though rare locally and largely unknown during the time of *Herschel*, these low-redshift galaxies are plausible analogs to early star-forming galaxies. Such local low-metallicity dwarf galaxies are natural extensions to the *SALTUS* low-redshift GTO efforts to measure and resolve the injection of feedback energy on small spatial scales far beyond the capabilities of *Herschel* [17] or a 2-m-class far-IR mission. With the same observables as the GTO program of star-forming spirals and starbursts, and the most commonly detected lines found with ALMA at high redshifts, *SALTUS* will enable ~200 pc-scale maps of the far-IR fine structure lines and underlying dust continuum in low-redshift analogs of the first generation of galaxies.

*5.6 Spectral Line Survey*

The benefits of performing high-frequency spectral line surveys were recognized and exploited by HIFI/Herschel. HIFI spectral surveys resulted in the first reported detections of SH+, HCl+, $H_2O$+, and $H_2Cl$+ [18, 19, 20, 21]. The formation of these small hydrides typically represents the first step in gas phase chemistry routes toward molecular complexity in space. However, the spectral surveys with HIFI were severely limited by sensitivity. *SALTUS's* increased aperture together with improved detector performance results in a **>30 times** increased sensitivity for these unresolved sources and can be expected to lead to an enhanced line density and discovery space.

*5.7 Broad and versatile follow-up capabilities for the 2030s*

Thanks largely to the advent of the large imaging and spectroscopic components of the Sloan Digital Sky Survey, modern extragalactic astrophysics is awash in observational campaigns that target objects pre-selected from multiwavelength surveys. This trend shows no sign of slowing: the first year of *JWST* operations and the ongoing community planning exercises for Rubin's *LSST* and *Roman* emphasize that key advances in our understanding are often only made possible by pointed follow-up programs. Another early lesson from *JWST*'s first year is that rapid advances are made possible when large windows in discovery space are opened: unexpected galaxy populations are found for the first time, objects that were once thought rare are revealed to be common, and flawed assumptions from previous galaxy evolution models are laid bare.

*SALTUS* is the only far-IR observatory capable of delivering on this clear need from the community and is the only facility that allows for genuine discovery space similar to *JWST*. Smaller apertures are not sufficiently sensitive and are plagued by spatial and spectral confusion that precludes the detection of faint objects without *a priori* knowledge of the position and redshift of every galaxy within a given field of view.

This science objective encapsulates the enormous variety of extragalactic science programs we expect from the community, assembled in much the same way as *HST* and *JWST* observing programs. To illustrate one example, *JWST* and *ALMA* observations have revealed an astonishing number of obscured or heavily reddened AGN at 5<z<10 [22, 23]. These spectroscopically confirmed AGN are at least ~30-100x more common than previously expected [24] and imply dramatic revisions to our understanding of early supermassive black hole growth. *SALTUS* is the only observatory sensitive enough to detect and measure the bolometric luminosities and black hole growth rates of this unexpected new population of reionization-era AGN. Even the longest wavelength band with J*WST*/MIRI (~24 microns) only probes up to rest-frame 3 microns at z~7, far bluer than the peak of AGN dust emission (~10-30 microns rest or ~70-200 microns observed-frame). Given recent number density measurements from *JWST*, we expect a single 10 hour pointing with *SALTUS* to detect the far-IR continua of ~10–30 obscured or heavily-reddened AGN at 5<z<10.

We stress that characterization of these unexpectedly common reionization-era AGN is illustrative: *SALTUS* is sufficiently sensitive and has the spatial resolution required to disentangle the far-IR light from galaxies selected in very heterogeneous surveys, allowing the observatory to be nimble enough to address not just the open science questions of the year 2023 but also the new questions raised in the 2030s. *SALTUS* is a versatile far-IR observatory with unmatched sensitivity, enabling genuine discovery space in extragalactic astronomy, capable of responding to current/future needs of the community.

Table 3 *SALTUS* GO Programs range from 77–83% available time depending on efficiency.

| SALTUS Investigation | GTO (hrs) Threshold | Observing Efficiency | | | GTO (hrs) Baseline | Observing Efficiency | | |
|---|---|---|---|---|---|---|---|---|
| | | 60% | 70% | 80% | | 60% | 70% | 80% |
| What is role of star formation in feedback in the Local Universe? | 250 | 5% | 4% | 4% | 550 | 2% | 2% | 2% |
| When did metals and dusts form in galaxies, affecting the process of star formation? | 100 | 2% | 2% | 1% | 500 | 2% | 2% | 1% |
| What are the roles of feeding black holes in galaxies from the early universe to today? | 300 | 6% | 5% | 4% | 1500 | 6% | 5% | 4% |
| Which feedback mechanism dominates as a function of time over cosmic history? | 100 | 2% | 2% | 1% | 1000 | 4% | 3% | 3% |
| How does the mass distribution in protoplanetary disks affect planet formation? | 250 | 5% | 4% | 4% | 1250 | 5% | 4% | 4% |
| What is the spatial distribution and evolution of water vapor and ice in protoplanetary disks? | 100 | 2% | 2% | 1% | 750 | 3% | 2% | 2% |
| How did Earth and Ocean Worlds get its water? | 200 | 4% | 3% | 3% | 500 | 2% | 2% | 1% |
| **Total SALTUS GTO Program (Hours)** | **1300** | 25% | 21% | 19% | **6050** | 23% | 20% | 17% |
| **Total GO Program (Percentage)** | | 75% | 79% | 81% | | 77% | 80% | 83% |

**6 Observing Allocations and Cadences**

**Table 3** shows the time allocated to achieve the *SALTUS* investigations for the Threshold (1 year timeframe) and Baseline (5 years) mission. As indicated in Table 3, even with a conservatively estimated observatory operational efficiency of 60%, *SALTUS* will meet the AO requirement of greater than 70% devoted to GO time. The observing efficiency depends on the location of targets, observatory settling time, and on-target observations before momentum unloading. In practice *SALTUS* will have an operational efficiency of between 60 – 80%, with the GO percentages in the Baseline program varying between 77-83% of available observing time.

With each *SALTUS* investigation requiring less than ~6% of available time, GO targets are available for observations so that GTO and GO programs can be supported near simultaneously.

The current best estimate for *SALTUS* operational efficiency is **83.5%,** based on detailed analysis of expected observatory performances. This is calculated with an average observational duration of 4-hours. *SALTUS* is capable of a 180° slew in 30 minutes. With a uniformly distributed set of targets over the sky, this gives a mean slew duration of 15 minutes and accounts for a 5.2% inefficiency. Momentum dumping adds an additional 2.1% inefficiency. Up/Downlinks take 4.2% overhead time and safe mode accounts for an additional 0.8% of out observing time. With an estimated damping coefficient of z=0.6%, and a review of similar booms performances, *SALTUS* damping values of 0.5% to 0.7%, gives settling times of 9 and 17 minutes respectively, giving an estimate required for settling time of 13 minutes, and efficiency impact of 4.2%. If needed, settling time can be shortened by increased damping design features and jerk-limiting attitude control algorithms. SALTUS' telecommunications architecture allows for continuous commanding, depending only on the availability of a ground station for command transmission. SALTUS' architecture does not limit time domain science cases that a GO might propose.

**7 Summary**

*SALTUS* has both the sensitivity and spatial resolution to address not just the open science questions of the year 2023, but, more importantly, the unknown questions that will be raised in the 2030s. *SALTUS is forward leaning and well suited to serving the current and future needs of the astronomical community.*

*Acknowledgments*

*Data Availability and Code*

This paper gives a mission concept in response to the NASA solicitation NNH23ZDA0210 for the 2023 Astrophysics Probe Explorer (APEX). The figures presented in the paper represents examples of predicted *SALTUS* observatory capabilities and does not derive from existing publicly accessible sources.

**Gordon Chin** is at NASA's Goddard Space Flight Center. He received his BA, MA, and PhD degrees from Columbia University. He has served as Project Scientist for the Lunar Reconnaissance Orbiter (*LRO*) and Submillimeter Wave Astronomy Satellite. He was a member of the NSF *ALMA* Management Advisory Committee and chair in 2003-2004. He was detailed to NASA HQ as Program Scientist for selection of the 2$^{nd}$ set of Small Explorer missions. He is a Co-I for the Lunar Exploration Neutron Detector on LRO. He is the PI and CoI for the NASA's DALI and MaTISSE instrument programs. He has authored about 200 publications.

**Carrie M. Anderson** is a research scientist at NASA Goddard Space Flight Center (GSFC). She received a BS in physics from Arizona State University in 2000, and MS and PhD degrees in Astronomy from New Mexico State University in 2003 and 2006, respectively. She is the author of more than 45 papers in refereed journals and has written one book chapter. Her research focuses on the remote sensing of planetary atmospheres, primarily in the areas of thermal structure and composition, using space- and ground-based data, in the visible, near-IR, mid-IR, far-IR, and submillimeter spectral regions. Her research also includes laboratory transmission spectroscopy measurements of ice films in a high-vacuum cryo-chamber located in her Spectroscopy for Planetary ICes Environments (SPICE) laboratory at NASA GSFC.

**Jennifer Bergner** is an Assistant Professor of Chemistry at UC Berkeley. She received her BS degree from University of Virginia and MA and PhD from Harvard in 2019. Her astrochemistry group uses a variety of tools to explore the chemistry at play in protostars and protoplanetary disks, the progenitors of planetary systems. With cryogenic vacuum experiments she mimics the extremely low temperatures and pressures of star-forming regions in the lab to explore the chemical and microphysical behavior of volatile ices. She also uses state-of-the-art telescope facilities like *ALMA* and *JWST* to observe the spectral fingerprints of volatile molecules in protostars and protoplanetary disks, providing insight into the chemical landscape of planet formation and the underlying physical processes which drive astrochemical evolution. She has about 128 publications.

**Nicolas Biver** holds a permanent research position at the CNRS (French center for scientific research) at Paris Observatory. He received his MS degrees in physics from Paris-Cité University in 1993, after graduating for the engineering high School "Ecole Centrale" in 1992. He defended his PhD in astrophysics on the observation and modelling of rotational lines of molecules in cometary comae from the Paris-Cité University in 1997. He is the author of more than 100 journal papers (25 as first author) and has written a book chapter on cometary chemistry (in Comets III, to be published in 2024). His current research interests include the molecular and isotopic composition of cometary volatiles and observations of solar system objects in the millimeter to submillimeter.

**Gordon L. Bjoraker** is a research scientist at the NASA Goddard Space Flight Center (GSFC). He received his BS in physics and astronomy from the University of Wisconsin in 1978 and his PhD degree in Planetary Science from the University of Arizona in 1985. He is the author of more than 90 journal papers. His research focuses on the remote sensing of planetary atmospheres, primarily in the areas of gas composition and cloud structure, using space- and ground-based data, in the near-IR, mid-IR, far-IR, and submillimeter spectral regions.

**Thibault Cavalie** is a permanent research scientist for the CNRS (French center for scientific research) at the Laboratoire d'Astrophysique de Bordeaux. He received his BS and MS degrees $_{59}$ in physics from the University of Bordeaux in 2003 and 2005, respectively, and his PhD in astrophysics on the observation oxygen species in the millimeter and submillimeter in planetary atmospheres in preparation for *Herschel* and ALMA from the University of Bordeaux in 2008. He is the author of more than 60 journal papers. His current research interests include mid-IR, far-IR and submillimeter spectroscopy using space- and ground-based data as well as chemical modeling of giant planet atmospheres to understand their formation and evolution.

**Michael A. DiSanti** is a research scientist at NASA Goddard Space Flight Center (GSFC). He received his BS and MS degrees in physics from the University of New Mexico in 1978 and 1980, respectively, and his PhD degree in physics from the University of Arizona in  He has coauthored more than 50 journal papers and two book chapters. His current research interests include the composition of cometary ices, and their relationship to solar system formation.

**Jian-Rong Gao** is a senior instrument scientist and head of Cryo-U Section within Instrument Science Group at SRON-Utrecht. He is also a part time faculty member of Quantum Nanoscience at TU Delft and is one of the theme leaders on space sensing at TU Delft Space Institute. He is a Co-I for NASA GUSTO balloon observatory. He is a co-chair of the Millimeter, Submillimeter, and Far-Infrared Detectors and Instrumentation for Astronomy conference, SPIE Astronomical Telescopes and Instrumentation, and also the international scientific advisory committee member of the 29th International Symposium on Space Terahertz Technology, ISSTT 2018, Pasadena, USA. Gao has published more than 250 papers in: THz phase gratings, wavefront, superconducting hot electron bolometer mixers and arrays, THz quantum cascade lasers, TES for FIR and X-ray, FDM, space instrumentations, KIDs, SIS mixers, nanostructures and physics.

**Leon Harding** is a senior staff engineer at Northrop Grumman Space Systems and is a mission architect in the Science & Robotic Exploration Systems group. He received a BSc (Hons) in experimental physics ('04), a MSc in astronomy ('08), and PhD in astrophysics with instrumentation ('12) all from the University of Galway in Ireland. Prior to Northrop Grumman, he held a research associate professorship at the Virginia Tech National Security Institute and was the assistant director of the Mission Systems Division where he led groups in space technology development applied to stellar/planetary magnetic activity, space weather, and planetary exploration. Before Virginia Tech, he was a Technologist at the Jet Propulsion Laboratory and a scientist at Caltech. He is the spacecraft lead systems engineer and deputy chief observatory architect for the SALTUS program.

**Paul Hartogh** leads the atmospheric science group in the planetary department of the Max Planck Institute for Solar System Research (MPS). He received his diploma and PhD in physics from the University of Gottingen in 1985 and 1989, respectively. He is the author of more than 250 journal papers and has written more than 10 book chapters. His current research interests include atmospheres in the solar system and as the Principal Investigator of the Submillimeter Wave Instrument (SWI) on the JUpiter ICy moons Explorer (JUICE) with a focus on the Jupiter system.

**Qing Hu** is a principal investigator in the Research Laboratory of Electronics (RLE) at the Massachusetts Institute of Technology (MIT). He received his B.A. from Lanzhow University in 1981 and his Ph.D. in physics from Harvard University in 1987. From 1987 to 1989, he was a postdoctoral associate at University of California, Berkeley. He joined the MIT faculty in 1990 in the Department of Electrical Engineering and Computer Science. He was promoted to full professor in 2002. Professor Hu has made significant contributions to physics and device applications over a broad electromagnetic spectrum from millimeter wave, THz, to infrared frequencies. Among those contributions, the most distinctive is his development of high-performance terahertz (THz) quantum cascade lasers (QCLs). He is a Fellow of the Optical Society of America (OSA), a Fellow of the American Physical Society (APS), a Fellow of the Institute of Electrical and Electronics Engineers (IEEE), and a Fellow of the American Association for the Advancement of Science (AAAS). He is the recipient of 2012 IEEE Photonics Society William Streifer Scientific Achievement Award. He has been an Associate Editor of Applied Physics Letters since 2006 and was the co-chair of 2006 International Workshop on Quantum Cascade Lasers. He has authored over 200 publications.

**Daewook Kim** is an associate professor of Optical Sciences and Astronomy at the University of Arizona. He has devoted his efforts to a multitude of space and ground-based large optical engineering projects. His primary research focuses on precision freeform optics design, fabrication, and various metrology topics, including interferometry and dynamic deflectometry. His contributions cover a broad spectrum of wavelengths, ranging from radio to x-ray. .Kim's and academic contributions include authoring over 300 journal/conference papers and serving as an associate editor for Optics Express. His academic achievements have led to his recognition as an SPIE Fellow, he was elected to the SPIE Board of Directors for the term spanning 2024 to 2026.

**Craig Kulesa** is an Associate Research Professor at Steward Observatory. He received a PhD in 2002 from the University of Arizona. Dr. Kulesa's main research area is the Galactic interstellar medium, with a special emphasis on the broad understanding of the life cycle of interstellar gas as it relates to star formation. Aspects of this evolutionary cycle include the formation and destruction of molecular clouds and the direct feedback mechanism between stars and gas. To study these processes, he is working on a variety of new infrared and submillimeter instrumentation. He is deputy-PI of both the 'Supercam' 64-beam imaging spectrometer for the SMT, and the balloon-borne Stratospheric Terahertz Observatory. He has also recently deployed a first-generation submillimeter telescope to the summit of the Antarctic plateau (HEAT) and is completing an infrared imager and echelle spectrometer for the MMT (ARIES). He is a CoI on NASA's GUSTO balloon observatory. He has authored about 150 publications.

**David T. Leisawitz** is an astrophysicist and Chief of the Science Proposal Support Office at NASA's Goddard Space Flight Center. He received a Ph.D. in Astronomy from the University of Texas at Austin in 1985. His primary research interests are star and planetary system formation, infrared astrophysics, wide-field spatio-spectral interferometry, and far-infrared space interferometry. He is NASA Center Study Scientist for the Far-IR Surveyor, Mission Scientist for the Wide-field Infrared Survey Explorer (WISE), and earlier served as Deputy Project Scientist for the Cosmic Background Explorer (COBE) under Project Scientist and mentor Dr. John Mather. He is Principal Investigator for "Wide-field Imaging Interferometry," a Co-Investigator on the "Balloon Experimental Twin Telescope for Infrared Interferometry (BETTII)," and member of a three-person External Advisory Panel for the "Far Infrared Space Interferometer Critical Assessment (FISICA)," a European Commission FP7 research program. In 2004-05, he served as Principal Investigator and science team lead for the Space Infrared Interferometric Telescope (SPIRIT) mission concept study. He has authored about 300 publications.

**Rebecca C. Levy** is a NSF Astronomy & Astrophysics Postdoctoral Fellow at the University of Arizona. She received her BS in Astronomy and Physics from the University of Arizona in 2015, and her MS and PhD degrees in Astronomy from the University of Maryland, College Park in 2017 and 2021, respectively. She has authored more than 40 refereed journal papers. Her research interests include studying extreme star formation, star clusters, stellar feedback, and the interstellar medium in nearby galaxies using multiwavelength ground- and space-based observations.

**Arthur Lichtenberger** is a Research Professor at the University of Virginia in Electrical and Computer Engineering and the NRAO Director of the UVA Microfabrication Laboratories (UVML). He received a BA from Amherst College in 1980 and M.S. and Ph.D. degrees from UVA in 1985 and 1987 respectively. He has built an internationally recognized research program in superconducting materials, devices, circuits and packaging for ultra-sensitive single pixel and array THz detectors, having collaborated with astronomical groups for the past 25 years to develop state of the art millimeter and submm wavelength mixers for use on radio telescopes throughout the world. His group's research includes the investigation of materials and microfabrication technologies for new terahertz devices, circuits and metrology. To date, he has been PI or Co-PI on over 25 million dollars of funding and an author on over 150 papers.

**Daniel P. Marrone** is an Associate Professor and experimental astrophysicist in the Department of Astronomy at the University of Arizona. He received his PhD at Harvard in 2006. His research addresses galaxies and cosmology and fundamental physics through a variety of observational tools. He is particularly interested in galaxy clusters and their cosmological applications, galaxy formation in the early universe, and the physics of the supermassive black hole in Sagittarius A*. He has developed new instruments, primarily at centimeter to submillimeter wavelengths. He has authored about 190 publications.

**Joan Najita** is an Astronomer at NSF's NOIRLab and its Head of Scientific Staff for User Support. She was formerly the Chief Scientist at the National Optical Astronomy Observatory (NOAO) and served on its scientific staff since 1998. In 1993 she received her PhD from University of California, Berkeley. Najita has been responsible for strategic planning, science career development, science communications, and the health of the scientific environment at the Observatory. Her interests include traditional research topics (such as star and planet formation, exoplanets, and the Milky Way), advocacy for the development of new research capabilities (such as infrared spectroscopy and massively multiplexed wide-field spectroscopy), as well as the sociological context of astronomy (such as the nature of discovery in astronomy, and its science sociology and resource allocation practices). She has a lifelong interest in communicating science to the public and in the role of science in society. Joan Najita has been named a 2021–2022 fellow at Harvard Radcliffe Institute, joining artists, scientists, scholars, and practitioners in a year of discovery and interdisciplinary exchange in Cambridge. She has authored about 190 publications.

**Dimitra Rigopoulou** is a Professor of Astrophysics at University of Oxford. She obtained a Master's in Astrophysics followed by a DPhil in Astrophysics from Queen Mary College, University of London. Her primary interests are on star-formation and galaxy evolution. She is interested in a class of galaxies called Luminous Infrared Galaxies. These objects are like `lighthouses' as they display the most extreme star-forming activity in the Universe. For her research she uses both imaging and spectroscopic techniques that allow investigations at sites of intense star-forming activity, the presence of an Active Galactic Nucleus (a black hole) and how the two phenomena co-exist and interact. She has authored about 80 publications.

**George H. Rieke** is the Regents Professor of Astronomy and Planetary Sciences at the University of Arizona in Tucson and the former Deputy Director of the Steward Observatory. He has a PhD from Harvard. He led the experiment design and development team for the Multiband Imaging Photometer for the Spitzer instrument on NASA's infrared Spitzer Space Telescope, and currently chairs the science team of the Mid-Infrared Instrument for the James Webb Space Telescope. Rieke has made contributions in: the starburst phenomenon, where a galaxy's properties are dominated by a very violent and short episode of star formation; the Galactic Center, the "local" prototype for active galactic nuclei; the origin of the infrared outputs of other active galactic nuclei; characterizing planetary debris disks, signposts for other planetary systems that cannot be detected in other ways; studying members of our own planetary system, among other topics. He was the Principal Investigator for the Multiband Infrared Photometer for Spitzer (MIPS).

**Peter Roelfsema** is a senior scientist/project manager at SRON Netherlands Institute for Space Research. He has been involved in several satellite projects, currently as PM for the Dutch Athena/X-IFU contribution, and before that as PI for SPICA's SAFARI Far-IR spectrometer and as lead of the international SPICA collaboration. He was PI and ad-interim PI for Herschel/HIFI, and in the early Herschel development phase, he was one of the lead system engineers developing the Herschel ground segment concept and operational systems. Before Herschel he led the ISO/SWS operations team in Villafranca/Spain and the SWS analysis software development team. He started his scientific career as a radio astronomer, utilizing the WSRT,

VLA and ATNF to study radio recombination lines of galactic HII regions and nearby active galaxies. With ISO and Herschel he did (Far)IR spectroscopic work on galactic HII regions, studying e.g. PAH properties and metal abundance variations in our galaxy. He has published over 150 papers in astronomical journals conference proceedings and supervised a number of PhD students.

**Nathan X. Roth** is an assistant professor at American University, conducting his research off-site at the NASA Goddard Space Flight Center. He received his BS and MS degrees in physics and astrophysics from the University of Missouri–St. Louis in 2014 and 2016, respectively, and his PhD degree in physics from the Missouri University of Science & Technology in 2019. He is the author of more than 40 journal papers and has written one book chapter. His current research interests include radio and infrared spectroscopy of solar system objects, including comets and asteroids, and applications to planetary defense.

**Kamber Schwartz** holds a postdoctoral position at the Max Planck Institute for Astronomy in Heidelberg. She was a NASA Sagan Postdoctoral Fellow at the Lunar and Planetary Laboratory at the University of Arizona. She received a PhD in Astronomy & Astrophysics at the University of Michigan in 2018. She studies the evolution of volatile gas during planet formation, with the goal of determining the amount of volatile carbon, nitrogen, and oxygen available to form planets. Her research combines observations from the infrared to the millimeter, using facilities such as *ALMA*, *NOEMA*, and *JWST*, with physical/chemical modeling to constrain the timescales and mechanisms of volatile reprocessing. She has authored over 100 publications.

**Yancy Shirley** is an Associate Professor at the Steward Observatory. He received a B.S., Astronomy & Physics and Applied Mathematics from the University of Arizona in 1997, and a Ph.D., Astronomy from University of Texas in 2002. He works on a variety of projects studying low-mass and high-mass star formation, the interstellar medium, and chemical evolution within our galaxy and nearby galaxies by combining observations with radiative transfer modeling. He specializes in radio and infrared imaging and spectroscopy utilizing single-dish radio telescopes, interferometers, and space-based observatories. He has authored over 200 publications.

**Justin Spilker** is an Assistant Professor at Texas A&M University. He is interested in the quenching of galaxies - the processes that prevent galaxies from forming new stars and to keep them from forming stars over long timescales. He uses radio/submillimeter interferometers like *ALMA* and the *VLA* for these studies. He has studied a sample of very dusty, highly star-forming galaxies detected by the South Pole Telescope that are magnified by a foreground galaxy through gravitational lensing. By virtue of their selection, this sample tends to lie at higher redshift than other samples observed with either Herschel or SCUBA/SCUBA-2. He has authored about 180 publications.

**Antony A. Stark** is a senior Astronomy at the Center for Astrophysics, Harvard & Smithsonian Astrophysical Observatory. He received a BS from Caltech in 1975 and a PhD at Princeton in 1989. He is a pioneer of Antarctic Astronomy and is a founder and designer of the South Pole Telescope (SPT), an instrument for observational cosmology. He is the PI and designer of the Parallel Imager for Southern Cosmology Observations (PISCO), a photometric camera on the Magellan Clay telescope for taking fast simultaneous G, R, I, and Z band images. He is *He is interested in* a member of the STO and GUSTO balloon-borne telescope teams for Milky Way and Magellanic Cloud TeraHertz spectroscopy surveys of the dominant cooling lines of the interstellar medium. He has authored about 350 publications.

**Floris van der Tak** is a Senior Scientist in the Astrophysics program of the Netherlands Institute for Space Research (SRON), where his research interests include astrochemistry, the habitability of exoplanets, the physics of the interstellar medium, star formation, molecular spectroscopy and radiative transfer. He received a PhD from Leiden University in 2000. He was the Project Scientist for the SPICA/SAFARI instrument. He has authored about 216 publications.

**Yuzuru Takashima** is a Professor of Optical Sciences at the Wyatt College of Optical Sciences at the University of Arizona. His research interests include: laser beam steering device, MEMS-LiDAR, automotive LiDAR, single-chip LiDAR; image steering and foveation, giga-pixel display, multiplexed display, near-to-eye display, multi-perspective 3D display, on-chip 3D display, holographic display; information optics: optical and holographic data storage; THz space optics, camera optics; digital micromirror device, MEMS phase light modulators, computer generated holograms, metrology, photonics device design and modeling. He received a BS from Kyoto University, a MS and PhD (2007) from Stanford University,. He is a Senior Member of SPIE and OSA. He has authored about 160 publications.

**Alexander Tielens** is a professor of astronomy in the Astronomy Department of the University of Maryland, College Park. He received his MS and PhD in astronomy from Leiden University in 1982. He has authored over 500 papers in refereed journals and has written two textbooks on the interstellar medium. His scientific interests center on the physics and chemistry of the interstellar medium particularly in regions of star and planet formation.

**David J. Willner** is a Senior Astrophysicist at the Smithsonian Astrophysical Observatory in the Radio and Geoastronomy Division at the Center for Astrophysics, Harvard & Smithsonian. His main research interests are circumstellar disks and the formation of planets, and the development of aperture synthesis techniques. Much of his science program makes use of radio, millimeter, and submillimeter interferometers, including the Submillimeter Array, *ALMA*, and the *VLA*. He received an A.B. in Physics from Princeton University and a Ph.D. in Astronomy from the University of California. He frequently lectures on imaging and deconvolution in radio astronomy. He has authored about 450 publications.

**Edward J. Wollack** is an Astrophysicist at NASA's Goddard Space Flight Center. He received a PhD in Physics (1994) at Princeton University and a BS in Physics (1987) at the University of Minnesota. His research concentrates on the development and use of precision imaging systems for astrophysics and observational cosmology. His main research focus is on the characterization of diffuse astrophysical backgrounds and understanding their implications for structures on the largest scales in the universe. In carrying out these efforts he has contributed to the design of a variety of novel sensor, guided wave, and optical systems for ground, sub-orbital, and space-borne applications. He has authored about 780 publications.

**Stephen Yates** is an instrument scientist at SRON, the Netherlands Institute for Space Research (since 2006). He received a Ph.D. from the University of Bristol (2003), followed by work at the CNRS-CRTBT (now Institut Neél) Grenoble on experimental low temperature techniques (2003-2006). He has over 17 years working on physics and applications of MKID detectors for THz astronomy with around 70 publications. Recent focus is on end-to-end full system design and characterization of MKID based (THz/FIR) instrumentation. One aspect is on the development and application of the phase and amplitude beam pattern technique to characterize optical interfaces and the optical performance of as-built instruments.

**Erick Young** is a Senior Science Advisor at Universities Space Research Association. He is a widely recognized authority on infrared astronomy and the former Science Mission Operations Director for SOFIA. He specializes in designing science instruments and has participated in many NASA's space infrared astronomy missions. He was responsible for developing the far-infrared detector arrays on the Spitzer Space Telescope's Multiband Imaging Photometer for Spitzer. As SOFIA Science Mission Operations Director, he manages the airborne observatory's equipment, instruments, support facilities, and infrastructure. He was also responsible for the overall scientific productivity of the facility, including the Guest Investigator program. He has about 385 publications.

**Christopher K. Walker** is a Professor of Astronomy, Optical Sciences, Electrical & Computer Engineering, Aerospace & Mechanical Engineering, and Applied Mathematics at the University of Arizona. He received his M.S.E.E. from Clemson University (1980), M.S.E.E. from Ohio State University (1981), and Ph.D. in Astronomy from the University of Arizona (1988). He has worked at TRW Aerospace and the Jet Propulsion Laboratory, was a Millikan Fellow in Physics at Caltech, and has been a faculty member at the University of Arizona since 199. He has made many contributions to advance the field of terahertz astronomy. He has supervised sixteen Ph.D. students, led numerous NASA and NSF projects, authored/coauthored 130+ papers, and published two textbooks: "Terahertz Astronomy" and "Investigating Life in the Universe".

**Caption List**

**Fig. 1** The SALTUS Observatory uses a radiatively cooled, inflatable 14-meter off-axis aperture with sensitive far-IR, high and moderate resolving power systems to open a new window on our Universe. See Sec 3 for details on the SALTUS observatory architecture.

**Fig. 2** Simulated terrestrial atmospheric transmission spectrum (black) demonstrating SALTUS*'s* far-IR spectral region, inaccessible from the ground, and outside of *JWST* (green) and *ALMA's* (magenta) operational wavelengths. The tunable HiRX Bands 1 – 4 with SAFARI-Lite (blue), target critically important wavelengths that are significantly or completely blocked from the ground; this includes low energy transitions of $H_2O$ and its isotopologues, and other species such as HD 1-0 and HD 2-1.

**Fig. 3** *SALTUS* complements the capabilities of JWST and ALMA and exceeds the achievable performances of the Herschel and SOFIA observatories by more than two orders of magnitude. It provides sensitivities comparable to the Origins Space Telescope flagship concept.

**Fig. 4** Simulated *SALTUS* image at 2.5" angular resolution (middle) of the [CII] 158 μm emission in NGC 6611 (Pillars of Creation) is similar to the JWST image (left) and compared to the SOFIA-created map (right). SAFARI-Lite can map this 10 arcmin2 region in 10 hours and simultaneously provide maps in all diagnostic lines of photo-dissociation regions (PDRs) and HII regions in our galaxy (Sec. D.4.4) and the local group (Sec. D.2.1.1), probing the physical environment produced by radiation feedback of massive stars and its link to stellar clusters and its molecular core.

**Fig. 5** The 3σ confusion limit curves in the far-IR as functions of wavelength for three different sized telescopes: a possible 1.8 m (blue), the 3.5 m *Herschel* (black), and the 14-m SALTUS (red).

The individual points indicate the best achieved sensitivity for *Herschel* and the projected sensitivity for SALTUS mapping a square arcminute after only 10 hours of observation, demonstrating how SALTUS greatly outperforms smaller aperture space observatories.

**Fig. 6** A simulated SALTUS 4.74 x 4.74 arcmin COSMOS deep field at 45, 70, and 160 µm wavelengths shown (left three vertical panels) with logarithmic stretch from 0.2 – 1.2 mJy/beam (45 µm), 0.3 -5.7 mJY/beam (70 µm), and 0.4- 15 mJy/beam (160 µm). To produce the composite image (right), all three maps were convolved to the SALTUS resolution at 160 µm and the 45, 70 and 160 µm maps were encoded as blue, green and red. SALTUS will observe fewer than 1 galaxy per beam to a fraction of a mJy even at 160 µm.

**Fig. 7** The spectral range of SALTUS over cosmic time. Schematic representation of the spectral energy distribution of a dusty $3\times10^{12}$ L$_\odot$ star forming galaxy with redshift. Lines important to the science case and PAH features are traced through redshift, and dominant cooling lines ([OI], [OIII], [CII]) are labeled. Out to z ~ 3, SAFARI-Lite probes the peak of the dust continuum and the bulk of the dust emission. Beyond z ~ 3, SAFARI-Lite takes over from JWST/MIRI to probe the red-shifted mid-IR PAH emission features. The yellow color-coded region indicates the wavelength range of SAFARI-Lite. The lower solid black curve is the detection limit for SAFARI-Lite at R = 300 for pointed observations (1 hour, 5σ). The lower long-dashed line approximates a detection limit for wide PAH features, which span many channels. The short-dashed line is the SAFARI-Lite detection limit in mapping mode (1arcmin$^2$ area mapped in 1-hr at 5σ).

**Fig. 8** *SALTUS* follows the water trail from molecular clouds to oceans. The habitability of planets is closely tied to the presence of H$_2$O, which is formed in the shielded interiors of molecular clouds, transported to planet forming disks where volatiles are further chemically processed before

becoming part of planetesimals and comets beyond the snow line. Planetesimals and comets then deliver these volatiles to terrestrial planets and ocean worlds. *SALTUS* is designed to probe this important journey using low lying rotational $H_2O$ lines that probe cold gas with HiRX and the icy grain reservoir through their phonon modes in emission with SAFARI-Lite while we expect GOs to probe the later stages.

**Fig. 9** Image of the Fomalhaut planetary debris system with Spitzer at 24 μm (left) and JWST at 25.5 μm (right). By analogy, an increase in telescope aperture from by a factor of seven (e.g. from 2 meters to 14 meters in the far infrared) yields a huge increase in information content of the images.

**Fig. 10** D/H ratios record the low temperature heritage of interstellar chemistry, modified by subsequent processing in planet forming disks. The high resolving power of *SALTUS* HiRX will reveal the time evolution of the D/H ratio in planet forming environments by measuring HDO and $H_2O$ in various objects, and HD in the Giant Planets. These measurements provide critical insight into the earliest stages of planetesimal formation, and the origin of Earth's water.

**Table 1** *SALTUS* Instruments performs observations over a wide wavelength range with both high resolution coherent and moderate resolution incoherent detectors at high spatial resolution.

**Table 2** *SALTUS* Science Objectives Address Relevant Decadal Themes and Questions.

**Table 3** *SALTUS* GO Programs range from 77–83% available time depending on efficiency.